\documentclass[reqno,12pt]{article}
\pdfoutput=1
\usepackage{ae}
\usepackage[T1]{fontenc}
\usepackage[ansinew]{inputenc}
\usepackage{mathrsfs}
\usepackage{amsmath}
\usepackage{amssymb}
\usepackage{graphicx}
\usepackage{color}
\definecolor{darkblue}{cmyk}{0.9,0.9,0,0}
\definecolor{carageen}{RGB}{0,0.55,0}
\usepackage[colorlinks=true,linkcolor=darkblue,citecolor=darkblue,urlcolor=darkblue]{hyperref}
\usepackage{epsfig}
\usepackage{wasysym}
\usepackage{MnSymbol}

\usepackage{amsmath,amsfonts,amssymb,amsbsy}
\usepackage{graphicx,color}
\usepackage{graphics}
\usepackage{epsfig}
\usepackage{verbatim}
\usepackage{stmaryrd}
\usepackage{epsf}

\usepackage{amsmath, amssymb,graphicx}
\usepackage{epsfig}
\usepackage{verbatim}
\usepackage{amsfonts}
\usepackage{ulem}
\usepackage{cite}

\usepackage{tikz}
\usetikzlibrary{arrows,shapes}
\usetikzlibrary{patterns,fadings}

\newcommand{\btp}{\begin{tikzpicture}[baseline=0pt,scale=0.9,line width=0.7pt]}
\newcommand{\btpp}{\begin{tikzpicture}[baseline=-5pt,scale=0.25,line width=0.7pt]}
\newcommand{\etp}{\end{tikzpicture}}

\def\bc{\begin{center}}
\def\ec{\end{center}}

\usepackage{epsf}

\newcommand{\be}{\begin{equation}}
\newcommand{\ee}{\end{equation}}
\newcommand{\ba}{\begin{eqnarray}}
\newcommand{\ea}{\end{eqnarray}}
\newcommand{\nn}{{\nonumber}}

\newcommand{\beaa}{\begin{eqnarray}}
\newcommand{\eeaa}{\end{eqnarray}}

\newcommand{\Tr}{\,{\rm Tr}}

\definecolor{carageen}{RGB}{0.1,0.7,0.1}

\newcommand{\la}[1]{\label{#1}}

\DeclareFontFamily{OT1}{pzc}{}
\DeclareFontShape{OT1}{pzc}{m}{it}{<-> s * [1.10] pzcmi7t}{}
\DeclareMathAlphabet{\mathpzc}{OT1}{pzc}{m}{it}

\def\({\left(}
\def\){\right)}
\def\[{\left[}
\def\]{\right]}

\def\<{\langle}
\def\>{\rangle}

\def\cO{{\cal O}}

\def\nref#1{(\ref{#1})}

\def\cusp{ {\rm cusp}}

\def\wlangle{ \llangle }
\def\wrangle{ \rrangle }

\newcommand{\cN}{{\cal N}}

\begin{document}

\thispagestyle{empty}

\renewcommand{\thefootnote}{\fnsymbol{footnote}}
\setcounter{footnote}{0}
\setcounter{figure}{0}
\begin{center}
$$$$
{\Large\textbf{\mathversion{bold}
 An exact formula for the radiation of a moving quark in ${\cal N}=4$ super Yang Mills}\par}

\vspace{1.0cm}

\textrm{Diego Correa$^a$, Johannes Henn$^b$, Juan Maldacena$^b$ and  Amit Sever$^{b,c}$ }
\\ \vspace{1.2cm}
\footnotesize{

\textit{$^{a}$  Instituto de F\'{\i}sica La Plata, Universidad Nacional de La Plata, \\ C.C. 67, 1900 La Plata, Argentina} \\
\texttt{} \\
\vspace{3mm}
\textit{$^{b}$ School of Natural Sciences,\\Institute for Advanced Study, Princeton, NJ 08540, USA.} \\
\texttt{} \\
\vspace{3mm}
\textit{$^c$
Perimeter Institute for Theoretical Physics\\ Waterloo,
Ontario N2J 2W9, Canada} \\
\texttt{}
\vspace{3mm}
}

\par\vspace{1.5cm}

\textbf{Abstract}\vspace{2mm}
\end{center}

\noindent

We derive an exact formula for the cusp anomalous dimension at small angles.
This is done by relating
the latter
to the computation of
certain 1/8 BPS Wilson loops
which was performed by supersymmetric localization.
This function of the coupling also determines the power emitted by a moving quark in
${\cal N}=4$ super Yang Mills, as well as the coefficient of the two point function of
the displacement operator on the Wilson loop.
By a similar method we compute the near BPS expansion of the generalized cusp
anomalous dimension.

\vspace*{\fill}

\setcounter{page}{1}
\renewcommand{\thefootnote}{\arabic{footnote}}
\setcounter{footnote}{0}

\newpage
\tableofcontents

\section{Introduction}

In this article we study some aspects of the cusp anomalous dimension of locally BPS Wilson loops
 in
${\cal N} =4$ super Yang Mills.
We consider the leading order expansion of $\Gamma_{cusp}$ at small angles,
\be \la{cuspsmallang}
 \Gamma_{cusp}(\phi) = - B(\lambda  ,N)  \phi^2  + o(\phi^4)
 \ee
  where we defined a
function which we call the ``Bremsstrahlung function" $B$. $B$ is a function of
the coupling. We compute it exactly at all values of the coupling and for all $N$, by
relating it to a computation that uses supersymmetric localization
 \cite{Erickson:2000af, Drukker:2000rr, Pestun:2007rz,Drukker:2006ga,Drukker:2006zk,Drukker:2007dw,Drukker:2007yx,Bassetto:2008yf,Pestun:2009nn,Giombi:2009ms,Giombi:2009ds}.
We  derive an exact expression for $B$ as
 \ba \label{bwilson}
  B &=&  { 1 \over 2 \pi^2} \lambda \partial_{\lambda}  \log  \langle W_\ocircle \rangle
  \\
  \langle W_\ocircle \rangle&=& { 1 \over N } L^1_{N-1}( - { \lambda \over 4 N } ) e^{ { \lambda \over 8 N } } ~,~~~~~~~~~~~~~\lambda = g^2_{YM} N  \la{lagres}
  \\
  B&=&   { 1 \over 4 \pi^2} { \sqrt{\lambda } I_2( \sqrt{\lambda} ) \over I_1( \sqrt{\lambda} ) }  + o( 1/N^2 ) \la{bplanar}
  \ea
% \comm{{\bf A} - Changed $\pi^4\to\pi^2$ in (4)}
  where $L$ is the modified Laguerre polynomial and $ W_\ocircle  $ is the 1/2 BPS circular Wilson loop computed in
 \cite{Erickson:2000af,Drukker:2000rr,Pestun:2007rz}. The last line gives the planar expression.
  The result is for the $U(N)$ theory. For $SU(N)$ we simply subtract the $U(1)$ contribution,
$B_{SU(N)} = B_{U(N)} - { \lambda \over 16 \pi^2 N^2 } $.
  This quantity $B$ also determines the energy emitted by a moving quark
  \be  \la{powerem}
  \Delta E  = { 2 \pi  B }  \int dt      ( \dot v)^2
  \ee
  in the small velocity limit. The result for any velocity can be obtained by
  performing a boost   and it is the same old formula
  that one has in electrodynamics,
   up to the replacement $ {2  e^2  \over 3 } \to2 \pi  B $, see \cite{Mikhailov:2003er} for a discussion at strong coupling. Its appearance
  in \nref{powerem} is what prompted us to call it the Bremsstrahlung function.
\begin{figure}[h]
\centering
\def\svgwidth{14cm}
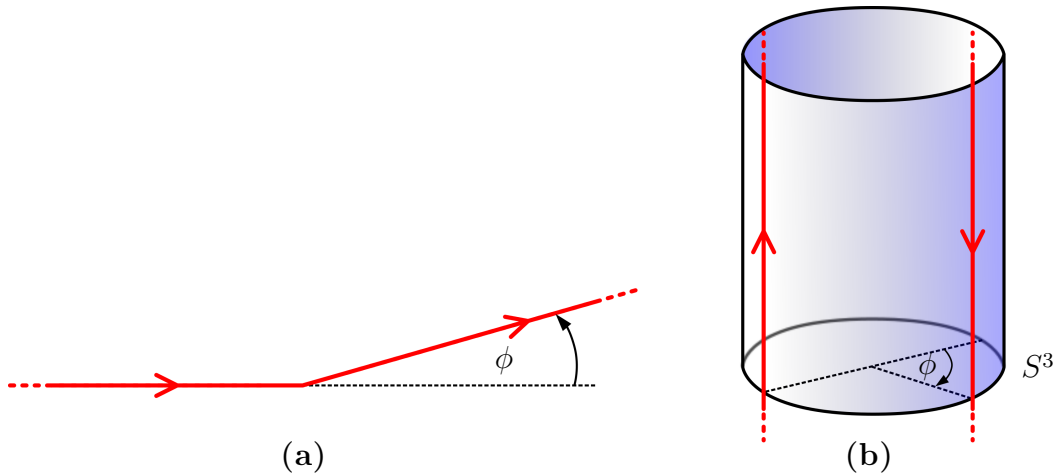
\caption{({\bf a}) A Wilson line that makes a turn by an angle $\phi$. ({\bf b}) Under the plane to cylinder map,  the same line is mapped to a quark anti-quark configuration. The quark and antiquark are sitting at two points  on $S^3$ at a relative angle of $\pi - \phi$. Of course, they are extended along the time direction.}\label{CuspDiagram}
\end{figure}

The cusp anomalous dimension is an interesting quantity that is related to a variety of
physical observables as particular cases.

Originally it was defined in \cite{PolyakovCusp} as the logarithmic divergence that arises
for a Wilson loop operator when there is a cusp in the contour. A cusp is a region where
a  straight line makes a sudden turn by an angle $\phi$, see figure \ref{CuspDiagram}(a).
In that case the Wilson loop develops a logarithmic divergence of the form
\be \la{cuspdef}
 \langle W \rangle \sim e^{ - \Gamma_{\cusp}(\phi,\lambda) \log { L \over \tilde \epsilon} }
\ee
 where $L$ is an IR cutoff and $\tilde \epsilon$ a UV cutoff.
 One can also consider the continuation $\phi = i \varphi$ so that now $\varphi$ is a boost
 angle in Lorentzian signature.

$\Gamma_{\cusp}$ is related to a variety of physical observables:

\begin{itemize}

\item
 It characterizes the IR divergences that arise when we scatter massive colored particles in the planar limit. Here
 $\varphi$ is the boost angle between two external massive particle lines. And for each consecutive pair of lines in the color ordered diagram we get a factor of \nref{cuspdef}, where
 $L$ is the IR cutoff and $\tilde \epsilon$ is set by the
 sum of the momenta of the two consecutive particles.  The angle is given by
 $ \cosh \varphi   =- { p_1\cdot p_2 \over \sqrt{ p_1^2 p_2^2 } } $.
 See e.g. \cite{Korchemsky:1991zp,Becher:2009kw} and references therein.

\item In $\cN =4$ super Yang Mills, the Regge limit of the four point massive
 scattering amplitudes on the Coulomb branch of $\cN=4$ SYM is governed by the
cusp anomalous dimension. As $t \gg s,m^2$, we have that {$\log \mathcal{A} \sim \log t  \; \Gamma_{\rm cusp}(\varphi)$} \cite{Henn:2010bk},
where $\mathcal{A}$ is the planar amplitude, divided by its value at tree level.

 \item
 The IR divergences of massless particles are characterized by $\Gamma^{\infty}_{\cusp}$ which
is the coefficient of the large $\varphi$ behavior  of the cusp anomalous dimension \cite{Korchemsky:1987wg,Korchemsky:1991zp}, $\Gamma_{\cusp} \sim \varphi\, \Gamma_{\cusp}^\infty $.   For ${\cal N}=4$ super Yang Mills,
$\Gamma^{\infty}_{\cusp}$ was computed in the seminal paper
 \cite{BES}.  Note that $\Gamma^{\infty}_{\cusp}$ is also sometimes called the ``cusp
 anomalous dimension'' though it is a particular limit of the general, angle dependent, ``cusp
 anomalous  dimension''  defined in \nref{cuspdef}.

 \item
 By the plane to cylinder map this quantity is identical with the energy of a static quark and
 anti-quark sitting on a spatial three sphere at an angle $ \pi - \phi$. See figure \ref{CuspDiagram}(b).
 In other words, it is the quark anti-quark potential on an $S^3$.

  \item
  As we already remarked, in the small angle limit it behaves as in \nref{cuspsmallang}, and
  it is related to the amount of power radiated by a moving quark. This function $B$
   also appears as the coefficient of the two point function of the displacement operator
  on a straight Wilson loop.
  In other words, $\wlangle   \mathbb{D}\,\mathbb{D} \wrangle   = 12 B$, where the
  double brackets denote the expectation values on the Wilson line.  This relation is
  completely general and is proven in section \ref{appDisplacement}. This two point function
of the displacement operator has appeared in discussions of wavy lines and the loop equation \cite{Polyakov:2000ti,Semenoff:2004qr}.

  \end{itemize}

In ${\cal N}=4$ super Yang Mills we can also introduce a second
angle at a cusp \cite{Drukker:1999zq}. This second angle is related to the fact that the locally supersymmetric
Wilson loop observable contains a coupling to a scalar. This coupling selects a direction $\vec n$, where $\vec n$ is a point on $S^5$. The Wilson loop operator is
given by
\be
 W \sim Tr[ P  e^{ i \oint A\cdot dx + \oint |dx| \vec n\cdot \vec \Phi } ]
  \ee
  where we wrote it in Euclidean signature. One can consider a loop with a constant
  direction $\vec n $, with $\partial_\tau \vec n =0$. When we have a cusp we could
  consider the possibility of changing the direction $\vec n$ by an angle $\theta$,
  $\cos \theta = \vec n \cdot\vec n'$, where $\vec n$ and $\vec n'$ are the directions
  before and after the cusp.
Thus, we have a cusp anomalous dimension, $\Gamma_{cusp}(\phi, \theta)$, which is a function of two angles $\phi$ and $\theta$.
The former is the obvious geometric angle and the latter is an internal angle.
 The same generalized cusp anomalous dimension $\Gamma_{cusp}(\phi, \theta)$ also characterizes the planar IR divergences that arise when scattering massive w-bosons on the Coulomb branch of ${\cal N}=4$ SYM. There, $\cos\theta=\vec n_1\cdot\vec n_2$ is the angle between the Coulomb branch expectation values $\<\vec\Phi\>=\text{diag}(m_1\vec n_1,m_2\vec n_2,\dots)$ associated to a pair of color adjacent external w-bosons $W_{1,i}$ and $W_{i,2}$. This generalized cusp
anomalous dimension was computed to leading and subleading order in weak and strong coupling in refs. \cite{Drukker:1999zq} and \cite{Forini:2010ek,Forini:2012rh}, respectively.

When $\theta = \phi$ (and also when $\theta = -\phi$) the configuration is supersymmetric
and therefore the cusp anomalous dimension vanishes.
In this case the Wilson loop is BPS. It is
a particular case of the 1/4 BPS loops considered in  \cite{Zarembo:2002an}.
We will show that the leading order term in the expansion of $\Gamma_{cusp}$ around
the supersymmetric value is
\be \la{DevBPS}
\Gamma_{cusp}(\phi,\theta)  = - ( \phi^2 -\theta^2 ) {  1  \over 1 - { \phi^2 \over \pi^2 } } B( \tilde \lambda ) + o((\phi^2  -\theta^2)^2) ~,~~~~~~~~~ \tilde \lambda = \lambda ( 1 - { \phi^2 \over \pi^2 } )
\ee
where $B$ is the same function appearing in \nref{bwilson}. For $\theta =0$ and small $\phi$ it reduces
to \nref{cuspsmallang}.

Finally, we should mention that in \cite{IntegrabilityPaper} a TBA system of integral
equations is derived for  $\Gamma_{cusp}(\phi,\theta)$. This is done via the standard approach
of integrability in the gauge/string duality, see \cite{Beisert:2010jr} for a review.
In particular, a limit of those integral equations computes
 the function $B$, see \cite{IntegrabilityPaper} for the details. Thus the cusp anomalous dimension at small angles
allows us to connect results computed using integrability with results computed using
localization. In \cite{IntegrabilityPaper}  $\Gamma_{cusp}$, and the function $B$,
is computed to three loops, matching
the expansion of the function $B$ we have in this paper.

\section{ Relating $\Gamma_{cusp}$ at small angles to the circular Wilson loop }

The derivation consists of the following steps.
First we consider a  1/4 BPS Wilson loop considered in \cite{Drukker:2006ga} (see
also \cite{Drukker:2006zk,Drukker:2007dw,Drukker:2007yx,Pestun:2009nn,Giombi:2009ms,Giombi:2009ds}).
It has  a parameter $\theta_0$.\footnote{Not to be confused with $\theta$ introduced above.}
 Taking the derivative with respect to $\theta_0$  we can
compute  the two point function of the scalar field on the Wilson loop.
This two point function of the scalar field also gives the expansion of the cusp
anomalous dimension around  $\phi =\theta =0$.

In this section we will set $\phi=0$ and we will expand in $\theta$. This is
equivalent to expanding in $\phi$, since $\Gamma_{cusp}$ vanishes for
BPS Wilson loops with
$|\theta| = |\phi|$.
Then for small values of $\phi $ and $\theta$ we expect
\be
\la{smallphiandtheta}
\Gamma_{cusp} = ( \theta^2 - \phi^2 )  B ~,~~~~~~ { \rm for}~~~~ \phi, ~ \theta \ll 1
\ee

\subsection{Relating derivatives of the Wilson loop to the scalar two point function}

In \cite{Drukker:2006ga},   a  1/4 BPS
circular Wilson loop with the following
value of the scalar profile was considered:
\be
  n_1 + i n_2 = \sin \theta_0\, e^{ i \tau }  ~,~~~~~~~~~n_3 = \cos\theta_0 ~,~~~~~~~~~
 \ee
 where $\tau$ is the coordinate along the circle (which we take to have radius one).
 Here $\vec n = (n_1,n_2,n_3,0,0,0)$  is  the profile for the scalar field in the Wilson loop,
 $W = \Tr P e^{ i \oint\! A + \oint\! d \tau\,  \vec n\cdot\vec \Phi } $.
 The spatial part is the same as before (a unit circle).
 For $\theta_0 =0$ we have the 1/2 BPS circular Wilson loop, whose expectation value is a non-trivial known function of $\lambda$ and $N$ \cite{Erickson:2000af, Drukker:2000rr}.
 For $\theta_0 = \pi/2$ we have the Zarembo 1/4
 BPS loop \cite{Zarembo:2002an} that has zero expectation value.

In \cite{Drukker:2006ga} it was conjectured that the Wilson loop expectation value would be given by
 the same expression as the circular one, but with a redefined coupling constant.
 This was further supported in \cite{Pestun:2009nn} using a localization
 argument\footnote{To complete the proof the determinant for the fluctuations in the
normal directions to the localization locus would be needed.} (see also \cite{Drukker:2006zk,Drukker:2007dw,Drukker:2007yx,Giombi:2009ms,Giombi:2009ds,Giombi:2009ek} for further discussion and checks).

 In other words, we have
 \be
 \langle W_{\theta_0}\>(\lambda)   = \langle W_\ocircle\>  ( \lambda'  )
 ~,~~~~~~~~~~\lambda' = \lambda \cos^2\theta_0
 \ee
 Now we expand this equation around $\theta_0 =0$. When $\theta_0$ is small $\lambda' \sim \lambda(1 - \theta_0^2 ) $ and we have
 \be \la{equ}
 { \langle W_{\theta_0}\> -  \langle W_{\theta_0 =0}\>
 \over \langle W_{\theta_0 =0}\> }
 \sim - \theta^2_0\, \lambda \partial_{\lambda} \log \langle W_\ocircle\>
\ee
Now the left hand side of \nref{equ} also has an  alternative expression in terms of the two
point function for a scalar field. Namely, we can write
\ba \la{ratiowil}
 && { \langle W_{\theta_0}\> -  \langle W_{\theta_0 =0}\> \over \langle W_{\theta_0 =0}\> } = \theta_0^2 I
 \\
 I\!\!\! & =&\!\!\! { 1 \over 2 } \int\limits_0^{2\pi }\!\! d\tau \int\limits_0^{2 \pi }\!\!   d\tau' \,\hat  n^i(\tau)  \hat n^j(\tau')\, \llangle  \Phi^i(\tau)
 \Phi^j(\tau')  \rrangle\nn \la{intcircle}
\ea
 where $\Phi^i$ are the scalars in the 1 and 2 directions and $\hat n^i$ are unit
 vectors in the $[12]$ plane.  The double brackets denote expectation values along the contour
\be
 \llangle  \cO(t_1) \cO(t_2)  \rrangle={\< Tr[P\, \cO(t_1) e^{\int_{t_2}^{t_1} i A\cdot dx + |dx|\vec n\cdot\vec \Phi } \cO(t_2)  e^{\int_{t_1}^{t_2} i A\cdot dx + |dx|\vec n\cdot\vec \Phi } ] \>\over\< Tr[ P e^{ i \oint A\cdot dx + \oint |dx| \vec n\cdot\vec \Phi }]\>}  \nn
\ee
where $\cO$ are in the adjoint representation and inserted along the loop.

  Any second order
 term from $n_3$ vanishes because it is a one point function, and the conformal symmetry of
 the 1/2 BPS line implies that one point functions are zero.
 We now use that
\be \la{twoptscal}
\wlangle   \Phi^i(\tau) \Phi^j(\tau') \wrangle   =  {   \gamma\,  \delta_{ij}  \over 2 [ 1 - \cos(\tau -\tau')] }
 \ee
 This just follows from a conformal transformation  of  the expectation value on the straight line\footnote{ The simplest way to derive this is to write the correlator,
\be
\la{propcov}
\langle \Phi \Phi \rangle = - { 1 \over 2 X \cdot X' }
\ee
  in
 projective
coordinates $X$, with $X^2 = X^+ X^- + \sum_{i=1}^4 X_i^2 =0$.  Then write these    in two gauges
\ba
 ( X^+,X^-, X^1 , X^2, X^3,X^4 )_{\rm straight} &=&  ( 1 , -t^2 , t,0,0,0 )
 \\
 ( X^+,X^-, X^1 , X^2, X^3,X^4 )_{\rm circle} &=&  ( 1 , -1 , \cos\tau , \sin \tau, 0,0 )
 \ea
  }
 \be
 \wlangle    \Phi(t) \Phi(0) \wrangle   = { \gamma \over t^2 }\quad\rightarrow\quad \wlangle   \Phi (\tau) \Phi (0) \wrangle    = {\gamma   \over 2[ 1 - \cos(\tau  )] }
\ee
We can now do the integral \nref{intcircle} as
\be \la{equtwo}
I = { \gamma } { \pi \over 2 }
 \int\limits_{\tilde \epsilon}^{2\pi-\tilde \epsilon}\!\! dt\, { \cos t \over[1- \cos(t)]  } =   {2 \gamma \pi  \over\tilde  \epsilon } - \pi^2 \gamma  = - \pi^2 \gamma
\ee
 where we discarded a power law UV divergence. Inserting this in \nref{ratiowil} and equating
   it to \nref{equ} we derive an expression for the coefficient $\gamma$
\be
   \la{gammaexpr}
   \gamma  = { 1 \over \pi^2 } \lambda \partial_\lambda \log \langle W_\ocircle\>
\ee

 \subsection{Relating the two point function of the scalar to $ { 1 \over 2 } \partial_\theta^2\, \Gamma_{cusp}$ }

 We now need to relate $\gamma$ to the second derivative of the cusp anomalous dimension.
 For that purpose we view the cusp as coming from the energy of two static quarks that are
 sitting at opposite points on $S^3$. When we vary the relative internal angle of
 these two quarks we get
\be \la{expans}
 \Gamma_{cusp} =\theta^2 I_c ~,~~~~~~~~ I_c =  -  { 1\over 2} \int\limits_{-\infty}^{\infty}\!  d\tau\,  \wlangle \Phi(\tau) \Phi(0) \wrangle
 \ee
 where now $\tau$ is the time in the $R\times S^3$ coordinates. We have set one of the operators
 at zero and  extracted an overall length of time factor.
 The minus sign comes from the relation between
 the partition function and the energy,  $Z = e^{ - \Gamma_{cusp} (\text{Length of Time}) } $.
 In \nref{expans} we have
 the correlator for two points on a Wilson loop along the same line in global coordinates.
 We transform from the plane to global coordinates\footnote{This is done by writing the projective coordinates in global coordinates
\ba \la{CYLCO}
 ( X^+,X^-, X^1 , X^2, X^3,X^4 )_{\rm global}& =&  ( e^\tau , -e^{-\tau} , 1,0, 0,0 )
\ea
}.
Then the integral in \nref{expans} is
\be \la{placyl}
I_c = - {\gamma \over 4 } \[ \int\limits_{-\infty}^{-\tilde \epsilon} + \int\limits_{\tilde \epsilon}^{\infty} \]  { d \tau  \over ( \cosh \tau - 1) }  = - { \gamma \over \tilde  \epsilon } +  { \gamma \over 2 }= { \gamma \over 2 }
\ee
where we again discarded a power law UV divergent term.
In conclusion, we find that
\be \la{cusptheta}
\Gamma_{cusp}(\phi=0,\theta)  =   \theta^2   { \gamma \over 2 } + o(\theta^4)
\ee
Of course, since $\Gamma_{cusp}$ vanishes for $\theta = \phi$, this also determines the
expansion of the cusp for small $\phi$, $\Gamma_{cusp} = - ( \phi^2 - \theta^2) { \gamma/2}$.
Inserting the value of $\gamma$ computed in \nref{gammaexpr}  we obtain
\be
 \Gamma_{cusp} = -  B (\phi^2 - \theta^2)  ~,~~~~~~~~~~~{\rm for} ~~~~\phi , ~\theta \ll 1 ~,~~~~~~~{\rm with} ~~~~
 B = { 1 \over 2 \pi^2 } \lambda \partial_{\lambda }
\langle W_\ocircle\>
\ee
where $\langle W_\ocircle\>$ is the 1/2 BPS circular Wilson loop  \nref{lagres}.

\begin{figure}[h]
\vspace{3.5cm}
\centering
\def\svgwidth{14cm}
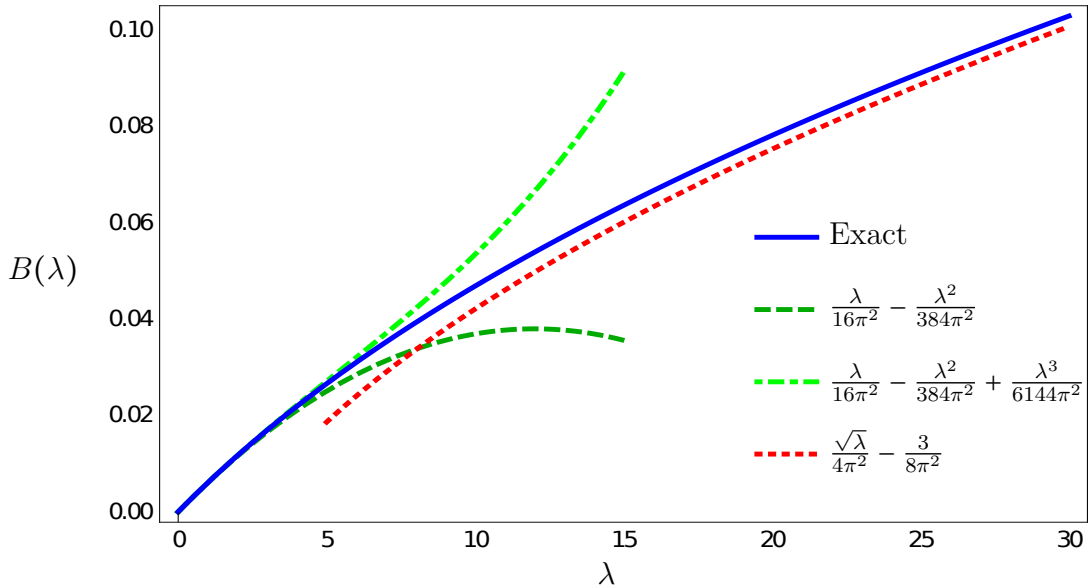
\vspace{-.5cm}
\caption{Plot of the Bremsstrahlung function $B$ in the planar limit (solid blue curve).
At weak coupling, the lower and upper dashed green curves denote the two- and three-loop approximation, respectively.
 It is interesting to note that the radius of convergence of the weak coupling
expansion is given by the first zero of $I_1$ in \nref{bplanar}, which is at $\lambda \sim  - 14.7$.
As one can see in the plot, the perturbative formulas become unreliable in that region. At the same time, we see that the first two orders of the strong coupling result (red dotted curve) give a qualitatively good approximation starting from that region.
%The first two orders at strong coupling are denoted by the red dotted curve.
}
\label{fig:bremsstrahlung_all_orders}
\end{figure}

We can expand this function at weak and strong coupling and, in the planar limit,
 we find
 \ba \la{weakBexp}
 B  & = &   \frac{\lambda }{16 \pi ^2}-\frac{\lambda ^2}{384 \pi ^2}+\frac{\lambda
   ^3}{6144 \pi ^2}-\frac{\lambda ^4}{92160 \pi ^2}+O\left(\lambda ^{5}\right)\\
B &  = &  \frac{\sqrt{\lambda }}{4 \pi ^2}\ -\ \frac{3}{8 \pi ^2}\ +\ \frac{3
  }{32 \pi ^2\sqrt\lambda}\ +\, \frac{3}{32 \pi ^2 \lambda
   }+O\left(  {\lambda } ^{-3/2}\right) \la{Bstrong}
 \ea
 The first two terms in each expansion agree with the results in \cite{Forini:2012rh}.
 The three loop term in the weak coupling expansion agrees with the explicit three loop computation in \cite{ThreeLoopPaper}. It also agrees with the three loop expansion of the TBA equations in
 \cite{IntegrabilityPaper}.
Figure \ref{fig:bremsstrahlung_all_orders} shows a plot of $B$ for $\lambda\in[0,30]$.

 \section{ Near BPS expansion of the generalized $\Gamma_{cusp}(\phi,\theta)$ }

 Now we turn our attention to the generalized cusp $\Gamma_{cusp}(\phi, \theta )$. When
$\phi = \pm \theta$,  it is zero since the configuration is BPS. Here we derive a simple
expression for the first order deviation away from the BPS value. Namely, we define a function $H$ as the
first term in the expansion away from the BPS limit at $\phi = \theta$,
\be
 \Gamma_{cusp} = - ( \phi - \theta ) H( \phi , \lambda )~,~~~~~~~~~ \theta -\phi \ll 1
 \ee
 We will show below that
\be\la{nearBPS}
 H( \phi, \lambda ) = { 2 \phi \over 1 - { \phi^2 \over \pi^2 }} B( \tilde \lambda ) ~,~~~~
 \tilde \lambda = \lambda ( 1 - { \phi^2 \over \pi^2 } )
 \ee
where $B$ is the same Bremsstrahlung function we had before  in \nref{bwilson}.

In order to derive this formula we need to consider a class of 1/8 BPS Wilson loops discussed in
 \cite{Drukker:2006zk,Drukker:2007dw,Drukker:2007yx,Pestun:2009nn,Giombi:2009ms,Giombi:2009ds,Giombi:2009ek}.
These are Wilson loops where the contour lives in  an $S^2$ subspace of $R^4$ or $S^4$.
These are BPS if the coupling to the scalars is chosen as follows.  We consider
a six dimensional vector of the form  $\vec n = ( \vec m  , 0,0,0) $ where
$\vec m$ is a three dimensional unit vector.
 If we call $\vec x$ the three dimensional unit vector
parametrizing the $S^2$, then we choose
\be
 \vec m =  \vec x \times \dot{ \vec x}  ~,~~~~~~~~(\vec x)^2 = ( \dot{ \vec x} )^2 =1, ~~~~~ \dot{ \vec x } = { d \vec x \over d t }
\ee

\begin{figure}[h]
\centering
\def\svgwidth{5cm}
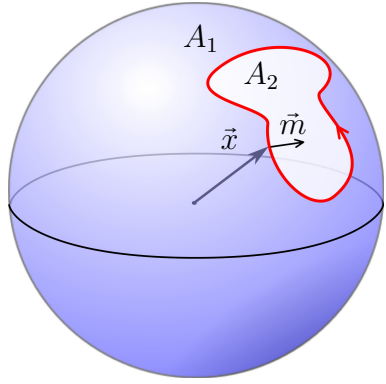
\caption{General class of 1/8 BPS Wilson loops. The loop lies on a two sphere $S^2$.
The vector that specifies the scalar couplings can be viewed as the unit vector on $R^3$ that
is orthogonal to the contour and lies on the two sphere, denoted here at a point by $\vec m$.
The two sphere is divided in two regions, one    with area   $A_1$ and  the other with area $A_2$.}\label{GeneralBPS}
\end{figure}
As conjectured  in
\cite{Drukker:2006zk,Drukker:2007dw,Drukker:2007yx}, and further argued and discussed in \cite{Pestun:2009nn,Giombi:2009ms,Giombi:2009ds,Giombi:2009ek},  the result for a non-intersecting Wilson loop of this kind is  given by the answer  for
the ordinary circular Wilson loop \nref{lagres}, but with $\lambda$ replaced by
\be \la{refpla}
\lambda \to \tilde \lambda = \lambda {   4 A_1 A_2 \over A^2 }  = \lambda 4 { A_1 \over A} { A_2 \over A}
\ee
where $A_1$ and $A_2$ are the areas of the two sides of the contour and $A=A_1 +A_2$ is the total area of
the two sphere.
\begin{figure}[h]
\centering
\def\svgwidth{11cm}
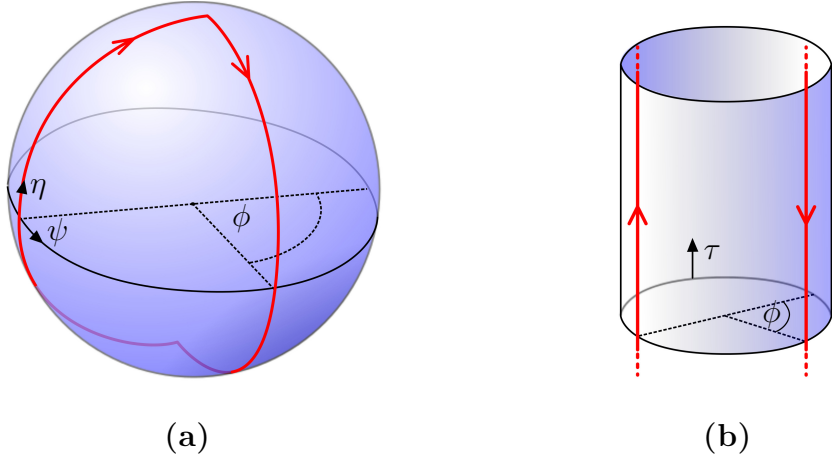
\caption{ ({\bf a}) Contour with two longitude lines separated by an angle $ \pi -\phi$. ({\bf b}) The
same contour on the cylinder.}\label{TwoLongitude}
\end{figure}

An interesting contour, considered in \cite{Drukker:2006zk,Drukker:2007dw,Drukker:2007yx,Giombi:2009ms},
consists of two longitude lines separated by an angle
$\pi - \phi$ and going all the way to the poles, see figure \ref{TwoLongitude}(a).
 In this case, \nref{refpla}, becomes
\be \la{tildelambdadef}
\tilde \lambda = {\lambda }\, 4  { ( \pi - \phi) \over  2 \pi }   { (\pi + \phi) \over  2 \pi }   = \lambda
( 1 - { \phi^2 \over \pi^2} )
\ee
The result of this Wilson loop is then
\be \langle W_\phi\> =
\langle W_\ocircle\>(\tilde \lambda )
\label{33}
\ee
 where $\langle W_\ocircle\>$ is the 1/2 BPS circular Wilson loop expectation value
given in \nref{lagres} and $\tilde \lambda$ is in \nref{tildelambdadef}. Here $\langle W_\phi\>$
denotes the BPS contour which is  made out of two longitude lines as depicted in figure \ref{TwoLongitude}(a).

Notice that this Wilson loop has a BPS cusp with angles $\theta = \phi$ at the two poles.
This loop is mapped to two straight lines separated by an angle $\pi-\phi$ in the cylinder, as depicted in figure \ref{TwoLongitude}(b). The fact that there is no logarithmic divergence is simply the statement that
$\Gamma_{cusp}(\theta = \phi)=0$\footnote{The expectation values for the same contour in the sphere and in the cylinder are different because the corresponding conformal map has a quantum anomaly.}.

It is useful to consider the conformal map between $S^4$ and $R \times S^3$ in global coordinates, which maps the north and south pole of an $S^2\subset S^4$ into $\tau = \pm \infty$, where $\tau $ is the $R$ coordinate. Under this map
the two longitude lines we mentioned above are mapped into two straight lines, extended along $R_\tau$ and sitting at a relative angle $\pi -\phi$ on the $S^3$. See figure \ref{TwoLongitude}.
\ba
ds^2\!\!\! &=&\!\!\! d\eta^2 + \cos^2 \eta\, d\Omega_3 = \cos^2 \eta \left( d\tau^2 + d\Omega_3^2 \right)\la{StoRS3}\\
 \sin \eta\!\!\!&=&\!\!\! \tanh \tau ~,~~~~~~~ \cos\eta = { 1 \over \cosh \tau } ~,~~~~~~
 { d \eta \over d \tau } = { 1 \over \cosh \tau } \la{mapcoord}
\ea
Here the $S^2\subset S^4$ is described by picking a great circle on the $S^3$ parametrized by $\psi$,
together with the direction $\eta$ of (\ref{StoRS3}). Then the vector $\vec x $ we
discussed above is parametrized as
\be
\vec x = ( \sin \eta , \cos \eta \cos \psi, \cos\eta  \sin \psi )
\ee
For longitude lines with $\dot\psi=0$ the internal vector $\vec m$ reads
\be \la{intvector}
\vec m = \vec x \times \dot { \vec x } = \dot\eta\,( 0 ,\sin \psi ,-\cos \psi ) ~,~~~~~\text{for}~~~ \dot \psi =0,\ |\dot\eta|=1
\ee

\begin{figure}[h]
\centering
\def\svgwidth{11cm}
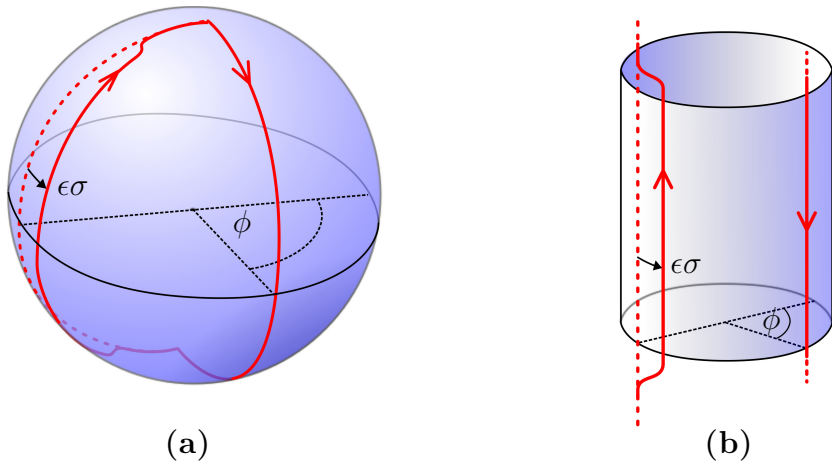
\caption{Contour we consider for the argument. We start from two longitude segments and we
deformed one of them. The deformed contour goes back to the old one near the poles. The deformation
is constant along most of the longitude line. In ({\bf a}) we see the diagram on the sphere. In ({\bf b}) we
see the diagram on the cylinder.}\label{DeformedLongitude}
\end{figure}

We start with a semicircle, or longitude line, at $\psi =0$ and another
at $\psi = \pi - \phi$. The internal vectors are
\be\la{mconf}
\vec m_{\psi=0} =( 0 ,0,-1)\ ,\qquad \vec m_{\psi=\pi-\phi} =( 0 ,-\sin\phi,-\cos\phi)
\ee
This configuration gives the supersymmetric cusp $\theta=\phi$.
We now consider a small deformation of this contour, which  will
help us in our arguments.
We deform the line that sits at $\psi =0$ in the following way
\be \la{smalldef}
\psi =  \epsilon\, \sigma(\eta )
\ee
where $\sigma(\eta)$ goes to zero at the two poles and it is
equal to one as soon as we get to a small distance from the poles.  See figure \ref{DeformedLongitude}.
Under the map into the straight lines \nref{mapcoord}, this gives a BPS deformation of the two
straight lines on $R\times S^3$. One of the lines is untouched. The line at $\psi =0$ is deformed
as in \nref{smalldef}.
  This deformation changes   the line so that it  goes to its   original value  at $\tau = \pm \infty$. For
a large region of values of $\tau$ the line and its internal angle are moved to a
 new location:  $\psi =  \epsilon$. Then there is a transition region that
  occurs at a large value of $\tau$.
   We will compute the change in the vacuum expectation value (VEV) of this Wilson loop in two ways and this will
tell us the VEVs of a scalar insertion and the displacement insertion,
$\Phi$ and $\mathbb{D}$,  on the straight lines.

The displacement operator displaces the contour in an orthogonal
direction.  Starting from a given contour we make an infinitesimal displacement $\delta x^j(t)$,  where $t$ is the proper time parameter along the contour. We consider displacements orthogonal to the contour, $\delta {\vec x}(t) \cdot \dot { \vec x}(t) =0$.  Then the displacement operator is   the functional derivative of the   Wilson loop operator  with respect to this displacement.
 In particular, we   write the infinitesimal change  as
\be \label{displde}
 \delta W =  P  \int\! d t\, \delta x^j(t)\mathbb{D}_j(t) W
 \ee
 where $\mathbb{D}_j$ is inserted at point $t$ along the contour. In other words $ P\mathbb{D}_j (t) W= P e^{i\int_{t}^\infty  A }\mathbb{D}_j (t) e^{ i\int_{-\infty}^t  A }$.

 For the locally BPS  Wilson loop in ${\cal N}=4$ super Yang Mills
  there is also an internal angle displacement operator which arises
 when we change $\vec n \to \vec n + \delta \vec n$.
 Since $\vec n$ has unit norm, $\vec n \cdot \delta
 \vec n  =0$. We can write then a small change of the Wilson loop as
 \be \la{smallwdi}
 \delta W = P \left[ \int\! dt\, \delta x^j\mathbb{D}_j\, W + \int\! dt\, \delta \vec n\cdot \vec \Theta\, W \right]
 \ee
 where the angle displacement operator is simply
 \be \la{smallind}
 \vec  \Theta = \vec \Phi
 \ee
  This operator is
 defined only for directions  orthogonal to
 the original angle $\vec n$ of the contour. Again the dimension
 of this operator is fixed by the symmetries. It has dimension one and it is in the same
 supermultiplet as the displacement \nref{disbps}  under the supergroup  preserved by the
 straight line, $OSp(4^*|4)$.   It is a BPS operator.
 An insertion of a scalar along the same internal direction as the original one ($\vec n \cdot \vec \Phi$)  is not protected
 \cite{Alday:2007he}.

Let us consider the case when $\psi$ varies in $\eta$, as in \nref{smalldef}. We get, to first order in $\epsilon$ 
\ba
\vec m  = \vec x \times \dot { \vec x } &=&(0,\epsilon\,\sigma,-1)+ \epsilon\,\partial_\eta \sigma \cos\eta\,  ( \cos \eta , -  \sin \eta , 0 )\nn
\\
&=&(0,\epsilon\,\sigma,-1)+ \epsilon\,\partial_\tau\sigma    ( { 1 \over \cosh \tau} , -  \tanh \tau , 0 )
\la{Mvalues}
\ea
where we wrote the result both in the sphere and cylinder coordinates using \nref{mapcoord}.
The crucial point here is the term proportional to $\partial_\tau \sigma $ in the second entry. Notice
that the term proportional to $\sigma$ in the second entry is what we expect from a constant change
in the internal angle position.
The source term  for the second scalar is  then
\be
 m_2 =  \epsilon [ \sigma(\tau ) - \tanh \tau \partial_\tau \sigma(\tau ) ]
\ee

We see that $\epsilon\, \sigma$ is the displacement in the angular direction, $\delta\phi =
\epsilon\, \sigma$. And, {\it when $\sigma $ is constant}, this is the same as the displacement
in the $\theta$ direction. However, when $\sigma $ varies, we get an extra source term for the
$\theta$ variation
\be
 \delta\phi =  \epsilon\, \sigma ~,~~~~~~~~~~~~~~ \delta \theta = \epsilon [ \sigma - \tanh \tau \partial_\tau \sigma ]
\ee
% Namely, the two vectors \nref{mconf} preserve an $SO(4) \subset SO(6)$ symmetry.
%note that the second line, the line that we did not modify, has an internal vector $\vec m_{\pi-\phi}$ given %by
%\nref{mconf} which is such that it has a zero component for the first scalar.

Now, for a general contour, if we make a small change in the contour $\delta x^i$ and a small
change in the internal orientation $\delta \vec n$, then we find that the change in the
expectation value of the Wilson loop is
\be  \la{wilsonchange}
{ \delta \langle W \rangle \over \langle W \rangle } = \int dt \left[ \wlangle\mathbb{D}_i \wrangle \delta x^i(t) +
\wlangle \vec  \Phi \wrangle \delta \vec n \right]
\ee
This is a general formula, valid for arbitrary displacements and changes in the internal orientation.
Note that $\delta  \vec n$ is orthogonal to $\vec n$ and thus the scalar fields in \nref{wilsonchange} are
scalars which are orthogonal to the original scalar appearing the loop at that point, which is $\vec n \cdot \vec \Phi$. Here $\mathbb{D}_i$ is the displacement operator in direction $i$.  See section  \ref{displacement} for further discussion
of this formula.
 We can now apply this general formula for the small change of the 1/8 BPS Wilson operator
we are considering.
We  find, to leading order in $\epsilon$,
\be \la{changecont}
    { \delta \langle W \rangle \over \langle W \rangle } =
 \epsilon \int d \tau  \left[  ( \wlangle\mathbb{D} \wrangle + \wlangle \Phi^2 \wrangle) \sigma(\tau)
-  \wlangle \Phi^2 \wrangle \tanh \tau\, \partial_\tau \sigma \right]
\ee
where $\Phi^2$ is the second scalar field.
We used that $\wlangle \Phi^1 \wrangle =0$,   due to a residual $SO(4) \subset SO(6)$
 symmetry of the configuration (\ref{mconf}).
The expectation values are the ones on the underformed contour. Since this contour has a
time translation symmetry, these expectation values are independent of the Euclidean time $\tau$. They
are simply constant and can be pulled out of the integral.
On the other hand, the same left hand side can be computed in terms of the general formula
\nref{refpla}. For the purposes of computing the areas in \nref{refpla} we can approximate
$\sigma =1$, since the change near the poles contributes in a negligible way to the area.
Thus the change in the expectation value is the change we have for the contour along the longitudes when
we change the angle by $\epsilon$
\be \la{changewB}
 { \delta \langle W \rangle \over \langle W \rangle } =  \epsilon\, \partial_\phi\log \langle W_\phi\> = \epsilon  { \partial \tilde \lambda \over \partial \phi}
\partial_{\tilde \lambda } \log[ \langle W_\ocircle \rangle(\tilde \lambda ) ] =-
 4 \epsilon   { \phi \over (1 - { \phi^2 \over \pi^2} ) }  B( \tilde \lambda )
\ee
Now this formula, combined with \nref{changecont} will allow us to compute the expectation values
of $\langle\mathbb{D} \rangle $ and $\langle \Phi^{2} \rangle $ as follows. First we note that the
BPS condition for $\delta \theta = \delta \phi$
implies that
\be
\wlangle\mathbb{D} \wrangle = -\wlangle \Phi^2 \wrangle
\ee
  Then  the only contribution to \nref{changecont} comes from the term with a derivative.
We can integrate it by parts and find
\ba
  { \delta   \langle W \rangle \over \langle W \rangle }
& = &\epsilon \wlangle \Phi^2 \wrangle \int d\tau { 1 \over \cosh^2 \tau } \sigma(\tau )
= 2 \epsilon   \wlangle \Phi^2 \wrangle   \label{changexp}
\ea
where we  used that $\sigma(\tau) =1$ in a large region around $\tau =0$ so that we can approximate
the integral of  $1/\cosh^2(\tau)$ (which is localized around $\tau = 0$) by its integral
 over $\int\limits_{-\infty}^\infty\!\! d\tau/\cosh^2(\tau) =2$.

In conclusion, by equating  \nref{changexp} and \nref{changewB} we have  managed to compute the expectation values
\be
 \wlangle\mathbb{D}\wrangle = -\wlangle \Phi^2 \wrangle =
2 { \phi \over (1 - { \phi^2 \over \pi^2} ) }  B( \tilde \lambda )
\ee

We now need to relate these expectation values to $\Gamma_{cusp}$. All we need is the
first order change of $\Gamma_{cusp}$ as we change the angular position $\phi$ from
its BPS value. This is given by the general formula \nref{wilsonchange}
where now we have constant change in $\delta \phi$  and $\delta \theta$.
We then get
\be
 \Gamma_{cusp} = -  \wlangle\mathbb{D} \wrangle \delta \phi  - \wlangle \Phi^2 \wrangle \delta \theta
 \ee
This is a general formula. If $\delta \theta = \delta \phi$, this vanishes as expected for a
BPS configuration.
It also implies, that expanding around $\theta = \phi$ we have
\ba
 \Gamma_{cusp} &=&  ( \phi - \theta) \wlangle \Phi^2 \wrangle  = -(\phi -\theta) H  ~,~~~~~~~\phi -\theta \ll 1\nn
\\
H &=& -\wlangle \Phi^2 \wrangle = { 2 \phi \over (1 - { \phi^2 \over \pi^2} ) }  B(  \tilde \lambda)
\la{finalh}
 \ea
which is the formula we wanted to derive \nref{DevBPS}. This is the same as \nref{DevBPS} because
for $\phi - \theta \ll \phi, \theta $ we have $\phi^2 - \theta^2 \sim 2 \phi (\phi - \theta )$.

In the weak coupling expansion we can check the first two loops with \cite{Forini:2012rh} and
the third loop with \cite{ThreeLoopPaper}.
One can also check that we get the correct answer at strong coupling from \cite{Forini:2012rh},
expanding around the $\theta -\phi$ limit. For that purpose, one should take the
formulas in appendix B of \cite{Forini:2012rh} which are written in terms of two parameters
 $q$ and $p$. We then  expand around $q=1$, we find that
$\Gamma_{cusp} = - { \sqrt{ \lambda } \over 2 \pi p } (\phi - \theta) $, to leading order in
 $\phi -\theta$. (Here $\phi -\theta$ is proportional to $(q-1)$). We also find that $p = { \pi \over \phi}   \sqrt{ 1 - { \phi^2 \over \pi^2 } }$, so that we have agreement with \nref{finalh}, and the strong coupling
form of $B$ in \nref{Bstrong}.

\section{Three related observables   }

In this section we would like to argue that the following three observables are
governed by the same function $B$.\begin{itemize}
\item
The derivative of the
cusp at zero angle
\be
B = \left. { 1 \over 2 } \partial_{\phi}^2\, \Gamma_{cusp}(\phi)\right|_{\phi =0}
\ee
\item
The two point function of the displacement operator on a straight Wilson loop
 \be
 \wlangle\mathbb{D}_i(t)\mathbb{D}_j(0) \wrangle = { \tilde \gamma\,  \delta_{ij} \over t^4 } ~,~~~~~~~~~~~~~\tilde \gamma = 12 B
 \ee
 \item
 The energy emitted by a moving quark, at small velocities,
\be
  \Delta E  = 2 \pi B  \int\! dt\, ( \dot v )^2
  \ee
\end{itemize}

The relations between these three observables are general for any conformal gauge
theory with a Wilson loop operator. In fact, they are valid for any line
defect operator, including 't Hooft line operators, etc.

We first start with  general remarks about the displacement operator and then
explain the relations.

%{\bf Note Added - }The same result for the power radiation $B$ (\ref{bwilson}) was obtained in \cite{Fiol:2012sg} by relating it to the correlation function of a Wilson line and the stress tensor \cite{Okuyama:2006jc}. That correlation function is another observable computed by function $B$.

\subsection{ The displacement operator for line defect operators}
\label{displacement}

In (\ref{displde}) we have written the infinitesimal change in a Wilson loop, due to 
a displacement orthogonal to the contour. The operator $\mathbb{D}$ is inserted along the contour and it is not a color singlet. It is a particular operator among a large class of operators that we can insert on a Wilson line
contour, see e.g. \cite{Dorn:1986dt,Polyakov:2000ti,Drukker:2006xg} for further examples.
In conformal theories these operators  are classified by their $SL(2) \times SU(2)$ quantum
numbers\footnote{$SL(2)\times SO(d-1)$ quantum numbers for a $d$-dimensional conformal theory.}.
Since the displacement $\delta x$ and the time $t$ appearing in \nref{displde}
have canonical dimensions, the dimension of $\mathbb{D}$ is protected and equal to two for all
values of the coupling. This is true for any line defect operator for any conformal field
theory.

For the ordinary Wilson loop operator, $ Tr[ P e^{ i \oint A } ] $,
the displacement operator at weak coupling is
given by
\be
\mathbb{D}_j =  i  F_{t j }
 \ee
 where $F_{tj}$ is the field strength and $t$ is the direction along the loop. (We wrote it
 in Euclidean signature).
 On the other hand if we are talking about the locally supersymmetric Wilson loop operator
 in ${\cal N} =4$ SYM,  which is $W =P e^{ i \int A + \int \vec n\cdot \vec \Phi }$, then the displacement operator
 is
 \be \la{disbps}
\mathbb{D}_j = i F_{t j } + \vec n \cdot \partial_j \vec \Phi 
 \ee
The coefficient of the two point function of the displacement operator on a straight line
\be   \la{displtwop}
\wlangle\mathbb{D}_i(t)\mathbb{D}_j(0) \wrangle = { \tilde \gamma\, \delta_{ij} \over t^4 }
\ee
is well defined since the normalization of $\mathbb{D}_i$ is fixed by its physical interpretation in
terms of the displacement. $\tilde \gamma > 0$ by unitarity (reflection positivity).
This number is an important characterization of the line defect operator\footnote{We thank D. Gaiotto for a discussion on this.}.

As an aside, the
two point function  \nref{displtwop} determines the expectation value of a Wilson loop with a wavy line profile
 as \cite{Polyakov:2000ti,Semenoff:2004qr}
\be
 \langle W \rangle = 1 + {\tilde \gamma \over 24  }
 \int\! dt \int\! dt'\,  {[ \dot { \vec \xi}(t) -  \dot { \vec \xi}(t')]^2 \over (t - t')^2 } + \cdots
 \ee
 Here $\vec \xi (t)$, with  $|\vec \xi| \ll 1, ~| \dot { \vec \xi } | \ll 1 $, is a small deviation from a straight line. We simply used that the contour is given by $ P e^{ \int\! dt\, \vec \xi(t)\cdot \vec{\mathbb{D}}(t) } W $, expanded in $\xi$, integrated by parts and dropped all power law UV divergent terms.  Again this is valid for any line defect operator in any CFT.

\subsection{The small angle $\Gamma_{cusp}$ and the two point function of the
 displacement operator }
 \label{appDisplacement}

Here we  relate  $B$,  defined as a second derivative of $\Gamma_{cusp}$, to $\tilde \gamma$,
  defined as the coefficient of  the two point function of the displacement operator \nref{displtwop}.

We consider the map between the plane ($R^4$) and the cylinder,  $R \times S^3$. This maps the
straight line into two lines along the $R$ direction and sitting at the ``north'' and ``south'' poles of
the $S^3$. For this configuration $\Gamma_{cusp}$ vanishes. The first derivative with respect to
$\phi$ also vanishes by the symmetries.
The second derivative with respect to $\phi$ can be computed in terms of the two point function of
the displacement operator evaluated in the cylinder coordinates
\be  \la{secondercu}
\Gamma_{cusp} \sim - \phi^2\, { 1 \over 2 } \int\! d\tau\,  \wlangle\mathbb{D}(\tau)\mathbb{D}(0) \wrangle =
- { \phi^2 \over 2 } \int\! d\tau  { \tilde \gamma \over 4 ( \cosh \tau -1)^2 }
\ee
Here we have used the same conformal change of coordinates that we discussed in  \nref{CYLCO},
except that now we have $\wlangle\mathbb{D}\,\mathbb{D} \wrangle = { 1 / ( 2 X \cdot X' )^2 } $ in projective coordinates.
In writing \nref{secondercu} we have neglected possible terms that involve one point functions, since
those would vanish by conformal symmetry.
We perform the integral in  \nref{secondercu} dropping its power law UV divergences to
  obtain
\be \la{btogamma}
 B = {\tilde \gamma \over 12 } ~~~~~~~~~~~~~~~~~{\rm with} ~~~~~ B = - { 1 \over 2 } \partial_\phi^2\, \Gamma_{cusp}(\phi) |_{\phi =0}
\ee
This formula relates $B$, defined as the second derivative of $\Gamma_{cusp} $ as in
\nref{cuspsmallang} and the two point function of the displacement operator \nref{displtwop}.

 \subsection{Relation between   $B$ and the  energy emitted by a moving
 quark}
\label{appRadiation}

In this section we relate the energy emitted by a moving quark to the two point
function \nref{displtwop}
of the displacement operator, which via \nref{btogamma} is related to the
cusp anomalous dimension at small angles.

The energy emitted by a moving quark, for small velocities can be written as
\be \la{smallv}
 \Delta E  = A  \int\! dt\, ( \dot v )^2
 \ee
We now relate $A$ to $\tilde \gamma $ as follows.
We consider a specific small displacement of the form $\delta x = \eta\, ( e^{ i \omega t } + e^{ - i\omega t })$, where $\eta$ is a small quantity.
The probability that the Wilson loop absorbs a quantum of energy $\omega$ is given in terms
of the two point function of the displacement operator, which couples to $\delta x$.
The absorption probability is
\ba \la{absorpro}
 p_{abs} &=&\parallel \eta\! \int\! dt\, e^{ - i \omega t }\mathbb{D}(t) |0 \rangle\!\parallel^2 =  T\,   |\eta|^2 \int\! dt\,  e^{ i \omega t }  \wlangle\mathbb{D}(t)\mathbb{D}(0) \wrangle   =
 \\
 &=&T  |\eta|^2  \tilde \gamma \int\limits_{-\infty}^\infty\! dt\, e^{ i \omega t } { 1 \over  ( t - i \epsilon)^4 }
  = { \pi \over 3 } |\eta|^2\, \tilde \gamma\, \omega^3\,  T
  \ea
 where $T$ is a long time cutoff.
 Here we consider the two point function in the explicit ordering stated, which results in the
 corresponding $i \epsilon$ prescription.
 To get the total absorbed energy we need to multiply \nref{absorpro} by a factor of $\omega$.
 Then we write $\omega^4 |\eta|^2$ as the integral  of the acceleration
 \be
  2 \omega^4 |\eta|^2 T =   \int\limits_0^T dt\, | \delta \ddot x |^2
 \ee
so that the emitted energy  is
 \be \la{deltae}
 \Delta E  = { \pi \over 6 } \tilde \gamma \int\! dt\,  ( \dot v )^2  = 2 \pi B \int\! dt\, (\dot v)^2
 \ee
 where we used \nref{btogamma}.
 This is again valid for any line defect operator in any conformal field theory.
  In the weak coupling limit it reduces to the  usual electrodynamics formula.
  The strong coupling form of $A$ was explicitly computed in
  \cite{Mikhailov:2003er}, which, of course, agrees with \nref{deltae} and the strong
   coupling form of $B$ \nref{Bstrong}. It was also show in  \cite{Mikhailov:2003er} that the
   energy emission formula is also the same (up to the overall coefficient)
   as in electrodynamics for any velocity. This follows from boost invariance.
     See \cite{Athanasiou:2010pv} for more discussion on properties of the
 radiation emitted by a moving quark at strong coupling in theories with a gravity dual.

\section{Conclusions}

In this paper we have derived an exact expression for the cusp anomalous dimension at small angles.
This corresponds to a small deviation from the straight line configuration. The result is written in eqs. \nref{bwilson} and \nref{lagres} and it is valid for all couplings and all $N$, in the $U(N)$ theory. The result was derived by relating the small deviation from the straight Wilson line
to the two point function of the displacement operator on a straight Wilson line.
The two point function of this displacement operator can then be obtained by considering the
1/4 (and 1/8) BPS family of Wilson loops discussed in \cite{Drukker:2006ga,Drukker:2006zk,Drukker:2007dw,Drukker:2007yx,Pestun:2009nn,Giombi:2009ms,Giombi:2009ds,Giombi:2009ek}. This family also contains the circular Wilson loop  but it can be computed exactly as a function
of the parameters that parametrize the family. The leading order deviation from the circular
Wilson loop is also given in terms of the two point function of the
displacement operator. This allows us to compute this two point function, using the
exact results conjectured in  \cite{Drukker:2006ga,Drukker:2006zk,Drukker:2007dw,Drukker:2007yx}
and proved in \cite{Pestun:2007rz,Pestun:2009nn}.

We have also emphasized that the small angle limit of the cusp is a function which
appears also in the two point function of the displacement operator of a Wilson loop
and in the formula for the energy radiated by a moving quark. These relations are general
and valid for any line defect operator in any CFT.

The answer in the planar limit is given by a ratio of Bessel functions \nref{bplanar}.
The function $I_2$
also appears in the expectation value of the operator ${\rm Tr}[Z^2]$ in the presence of a Wilson loop
 \cite{Semenoff:2001xp,Pestun:2007rz}.
This operator is in the same supermultiplet as the lagrangian. So it is likely that
this explains why this ratio is equal to the derivative of the loop with respect to the coupling.
Since the lagrangian is the same supermultiplet as the stress tensor, and a Ward identity relates the
stress tensor to the displacement operator  \cite{Gaiotto}, it is likely that
this could also explain the relations derived in this paper in an alternative fashion.

A very similar ratio of Bessel functions was recently conjectured in \cite{Basso:2011rs} to be related to the
first derivative with respect to the spin at zero spin for twist $J$ operators. The formula
in  \cite{Basso:2011rs} says that
\be
 \lim_{S \to 0 } {  \Delta(S) - J   \over S }  = { \sqrt{ \lambda } \over J } { I'_{J}( \sqrt{\lambda} ) \over I_{J}(\sqrt{\lambda } ) }
 \ee
 Of course, one needs to analytically continue in the spin $S$ to be able to write this formula.
 This looks similar to \nref{bplanar}, except that we would have to set $J=1$. But there
 are no closed string operators with $J=1$!.  On a Wilson loop there are twist one
 operators, so maybe there is a connection between the function $B$ and this small spin
 limit. But we leave this problem to the future.

Basically the same idea can be used to compute the leading deviation from the BPS limit
for the generalized cusp anomalous dimension $\Gamma_{cusp}(\phi , \theta)$. The leading
deviation is the coefficient of the term proportional to $\phi -\theta$. This is a non-trivial,
but simple, function of the angle (\ref{nearBPS}). Expanded in perturbation theory, this function of the angle
has the form $ \lambda^L \phi ( \phi^2 - \pi^2)^{ L-1}$, where $L$ is the loop order.
Note that the cusp anomalous dimension is intimately related to amplitudes. As we discussed
in the introduction, it governs the IR divergences for the scattering of massive particles in
the presence of massless gluons. As such, it is a  non-trivial function of the angle with a
given transcendentality. In ${\cal N}=4$ SYM, we have a total transcendentality $2L$ for an $L$
loop diagram. The $\log \mu_{IR}$ soaks up one unit of transcendentality. Thus $\Gamma_{cusp}(\phi)$ has transcendentality $ 2L -1$ at $L$ loops, which is precisely what we
see in this formula. For counting the transcendentality, we should assign $\phi = \log e^\phi$
transcendentality one. While a scattering amplitude is a function of many variables, $\Gamma_{cusp}$ is just a function of one variable $\phi$. We see that in this limit it is
a particularly simple function, which can be computed at all loop orders!.

As will be shown in \cite{IntegrabilityPaper} the same function $B$ can be computed
from an integral equation of the TBA type. This then connects the traditional integrability
approach with localization.  A similar connection would be more exciting for the ABJM theory
since it would allows us to compute the undetermined function of $\lambda$ which is present
in the integrability approach \cite{Nishioka:2008gz,Gaiotto:2008cg,Grignani:2008is,Gromov:2008qe}.

{\bf Note Added - } Shortly after this paper appeared, the same result for the power radiation $B$ (\ref{bwilson}) was obtained in \cite{Fiol:2012sg} by relating it to the correlation function of a Wilson line and the stress tensor \cite{Okuyama:2006jc}. That correlation function is another observable computed by function $B$.

{\bf Acknowledgements }

We would like to thank N. Arkani-Hamed, B. Basso, S. Caron-Huot, N. Drukker, D. Gaiotto, I. Klebanov,  V. Pestun and P. Vieira for discussions.

A. S. would like to thank Nordita for warm hospitality. This work was supported in part by   U.S.~Department of Energy grant \#DE-FG02-90ER40542.
Research at the Perimeter Institute is supported in part by the Government of Canada through NSERC and by the Province of Ontario through MRI.  The research of  A.S.
 has been supported in part by the Province of Ontario through ERA grant ER 06-02-293.
D.C would like to thanks IAS for hospitality. The research of D.C has been supported in part by a CONICET-Fulbright fellowship and grant PICT 2010-0724.


\begin{thebibliography}{99}



\bibitem{Erickson:2000af}
  J.~K.~Erickson, G.~W.~Semenoff and K.~Zarembo,
  ``Wilson loops in N=4 supersymmetric Yang-Mills theory,''
  Nucl.\ Phys.\ B {\bf 582}, 155 (2000)
  [hep-th/0003055].
  %%CITATION = HEP-TH/0003055;%%%\cite{Drukker:2000rr}



\bibitem{Drukker:2006ga}
  N.~Drukker,
  ``1/4 BPS circular loops, unstable world-sheet instantons and the matrix model,''
  JHEP {\bf 0609}, 004 (2006)
  [hep-th/0605151].
  %%CITATION = HEP-TH/0605151;%%


\bibitem{Drukker:2006zk}
  N.~Drukker, S.~Giombi, R.~Ricci and D.~Trancanelli,
  ``On the D3-brane description of some 1/4 BPS Wilson loops,''
  JHEP {\bf 0704}, 008 (2007)
  [hep-th/0612168].
  %%CITATION = HEP-TH/0612168;%%


\bibitem{Drukker:2007dw}
  N.~Drukker, S.~Giombi, R.~Ricci and D.~Trancanelli,
  ``More supersymmetric Wilson loops,''
  Phys.\ Rev.\ D {\bf 76}, 107703 (2007)
  [arXiv:0704.2237 [hep-th]].
  %%CITATION = ARXIV:0704.2237;%%


\bibitem{Drukker:2007yx}
  N.~Drukker, S.~Giombi, R.~Ricci and D.~Trancanelli,
  ``Wilson loops: From four-dimensional SYM to two-dimensional YM,''
  Phys.\ Rev.\ D {\bf 77}, 047901 (2008)
  [arXiv:0707.2699 [hep-th]].
  %%CITATION = ARXIV:0707.2699;%%

%\cite{Bassetto:2008yf}

\bibitem{Bassetto:2008yf}
  A.~Bassetto, L.~Griguolo, F.~Pucci and D.~Seminara,
  ``Supersymmetric Wilson loops at two loops,''
  JHEP {\bf 0806}, 083 (2008)
  [arXiv:0804.3973 [hep-th]]. $\bullet$ D.~Young,
  ``BPS Wilson Loops on S**2 at Higher Loops,''
  JHEP {\bf 0805}, 077 (2008)
  [arXiv:0804.4098 [hep-th]].
  %%CITATION = ARXIV:0804.4098;%%
  %%CITATION = ARXIV:0804.3973;%%


\bibitem{Pestun:2009nn}
  V.~Pestun,
  ``Localization of the four-dimensional N=4 SYM to a two-sphere and 1/8 BPS Wilson loops,''
  arXiv:0906.0638 [hep-th].
  %%CITATION = ARXIV:0906.0638;%%


\bibitem{Giombi:2009ms}
  S.~Giombi, V.~Pestun and R.~Ricci,
  ``Notes on supersymmetric Wilson loops on a two-sphere,''
  JHEP {\bf 1007}, 088 (2010)
  [arXiv:0905.0665 [hep-th]].
  %%CITATION = ARXIV:0905.0665;%%


\bibitem{Giombi:2009ds}
  S.~Giombi and V.~Pestun,
  ``Correlators of local operators and 1/8 BPS Wilson loops on S**2 from 2d YM and matrix models,''
  JHEP {\bf 1010}, 033 (2010)
  [arXiv:0906.1572 [hep-th]].
  %%CITATION = ARXIV:0906.1572;%%


\bibitem{Drukker:2000rr}
  N.~Drukker and D.~J.~Gross,
  ``An Exact prediction of N=4 SUSYM theory for string theory,''
  J.\ Math.\ Phys.\  {\bf 42}, 2896 (2001)
  [hep-th/0010274].
  %%CITATION = HEP-TH/0010274;%%


\bibitem{Pestun:2007rz}
  V.~Pestun,
  ``Localization of gauge theory on a four-sphere and supersymmetric Wilson loops,''
  arXiv:0712.2824 [hep-th].
  %%CITATION = ARXIV:0712.2824;%%


\bibitem{Mikhailov:2003er}
  A.~Mikhailov,
  ``Nonlinear waves in AdS / CFT correspondence,''
  hep-th/0305196.
  %%CITATION = HEP-TH/0305196;%%


\bibitem{PolyakovCusp}
  A.~M.~Polyakov,
  ``Gauge Fields as Rings of Glue,''
  Nucl.\ Phys.\ B {\bf 164}, 171 (1980).
  %%CITATION = NUPHA,B164,171;%%

 %\cite{Korchemsky:1991zp}

\bibitem{Korchemsky:1991zp}
  G.~P.~Korchemsky and A.~V.~Radyushkin,
  ``Infrared factorization, Wilson lines and the heavy quark limit,''
  Phys.\ Lett.\ B {\bf 279} (1992) 359
  [hep-ph/9203222].
  %%CITATION = HEP-PH/9203222;%%

%\cite{Becher:2009kw}

\bibitem{Becher:2009kw}
  T.~Becher and M.~Neubert,
  ``Infrared singularities of QCD amplitudes with massive partons,''
  Phys.\ Rev.\ D {\bf 79} (2009) 125004
   [Erratum-ibid.\ D {\bf 80} (2009) 109901]
  [arXiv:0904.1021 [hep-ph]].
  %%CITATION = ARXIV:0904.1021;%%

%\cite{Henn:2010bk}

\bibitem{Henn:2010bk}
  J.~M.~Henn, S.~G.~Naculich, H.~J.~Schnitzer and M.~Spradlin,
  ``Higgs-regularized three-loop four-gluon amplitude in N=4 SYM: exponentiation and Regge limits,''
  JHEP {\bf 1004} (2010) 038
  [arXiv:1001.1358 [hep-th]].
  %%CITATION = ARXIV:1001.1358;%%

  %\cite{Korchemsky:1987wg}

\bibitem{Korchemsky:1987wg}
  G.~P.~Korchemsky and A.~V.~Radyushkin,
  ``Renormalization of the Wilson Loops Beyond the Leading Order,''
  Nucl.\ Phys.\ B {\bf 283} (1987) 342.
  %%CITATION = NUPHA,B283,342;%%


\bibitem{BES}
  N.~Beisert, B.~Eden and M.~Staudacher,
  ``Transcendentality and Crossing,''
  J.\ Stat.\ Mech.\  {\bf 0701}, P01021 (2007)
  [hep-th/0610251].
  %%CITATION = HEP-TH/0610251;%%



\bibitem{Polyakov:2000ti}
  A.~M.~Polyakov and V.~S.~Rychkov,
  ``Gauge field strings duality and the loop equation,''
  Nucl.\ Phys.\ B {\bf 581}, 116 (2000)
  [hep-th/0002106].
  %%CITATION = HEP-TH/0002106;%%


\bibitem{Semenoff:2004qr}
  G.~W.~Semenoff and D.~Young,
  ``Wavy Wilson line and AdS / CFT,''
  Int.\ J.\ Mod.\ Phys.\ A {\bf 20}, 2833 (2005)
  [hep-th/0405288].


\bibitem{Drukker:1999zq}
  N.~Drukker, D.~J.~Gross and H.~Ooguri,
  ``Wilson loops and minimal surfaces,''
  Phys.\ Rev.\ D {\bf 60}, 125006 (1999)
  [hep-th/9904191].
  %%CITATION = HEP-TH/9904191;%%

%\cite{Forini:2010ek}

\bibitem{Forini:2010ek}
  V.~Forini,
  ``Quark-antiquark potential in AdS at one loop,''
  JHEP {\bf 1011}, 079 (2010)
  [arXiv:1009.3939 [hep-th]].
  %%CITATION = ARXIV:1009.3939;%%


\bibitem{Forini:2012rh}
  N.~Drukker and V.~Forini,
  ``Generalized quark-antiquark potential at weak and strong coupling,''
  JHEP {\bf 1106}, 131 (2011)
  [arXiv:1105.5144 [hep-th]].
  %%CITATION = ARXIV:1105.5144;%%



\bibitem{Zarembo:2002an}
  K.~Zarembo,
  ``Supersymmetric Wilson loops,''
  Nucl.\ Phys.\ B {\bf 643}, 157 (2002)
  [hep-th/0205160].
  %%CITATION = HEP-TH/0205160;%%


\bibitem{IntegrabilityPaper}
D.~Correa, J.~Maldacena and A.~Sever,
  ``The quark anti-quark potential and the cusp anomalous dimension from a TBA equation,''
  arXiv:1203.1913 [hep-th].

\bibitem{Beisert:2010jr}
  N.~Beisert, C.~Ahn, L.~F.~Alday, Z.~Bajnok, J.~M.~Drummond, L.~Freyhult, N.~Gromov and R.~A.~Janik {\it et al.},
  ``Review of AdS/CFT Integrability: An Overview,''
  Lett.\ Math.\ Phys.\  {\bf 99}, 3 (2012)
  [arXiv:1012.3982 [hep-th]].
  %%CITATION = ARXIV:1012.3982;%%


\bibitem{Giombi:2009ek}
  S.~Giombi and V.~Pestun,
  ``The 1/2 BPS 't Hooft loops in N=4 SYM as instantons in 2d Yang-Mills,''
  arXiv:0909.4272 [hep-th].
  %%CITATION = ARXIV:0909.4272;%%


\bibitem{ThreeLoopPaper}
 D.~Correa, J.~Henn, J.~Maldacena and A.~Sever,
  ``The cusp anomalous dimension at three loops and beyond,''
  JHEP {\bf 1205}, 098 (2012)
  [arXiv:1203.1019 [hep-th]].

\bibitem{Alday:2007he}
  L.~F.~Alday and J.~Maldacena,
  ``Comments on gluon scattering amplitudes via AdS/CFT,''
  JHEP {\bf 0711}, 068 (2007)
  [arXiv:0710.1060 [hep-th]].
  %%CITATION = ARXIV:0710.1060;%%AldayMaldacena}


\bibitem{Fiol:2012sg} 
  B.~Fiol, B.~Garolera and A.~Lewkowycz,
  ``Exact results for static and radiative fields of a quark in N=4 super Yang-Mills,''
  arXiv:1202.5292 [hep-th].
  %%CITATION = ARXIV:1202.5292;%%
  
%\cite{Okuyama:2006jc}

\bibitem{Okuyama:2006jc} 
  K.~Okuyama and G.~W.~Semenoff,
  ``Wilson loops in N=4 SYM and fermion droplets,''
  JHEP {\bf 0606}, 057 (2006)
  [hep-th/0604209].
  %%CITATION = HEP-TH/0604209;%%




\bibitem{Dorn:1986dt}
  H.~Dorn,
  ``Renormalization Of Path Ordered Phase Factors And Related Hadron Operators In Gauge Field Theories,''
  Fortsch.\ Phys.\  {\bf 34}, 11 (1986).
  %%CITATION = FPYKA,34,11;%%


\bibitem{Drukker:2006xg}
  N.~Drukker and S.~Kawamoto,
  ``Small deformations of supersymmetric Wilson loops and open spin-chains,''
  JHEP {\bf 0607}, 024 (2006)
  [hep-th/0604124].
  %%CITATION = HEP-TH/0604124;%% %\cite{Mikhailov:2003er}


\bibitem{Athanasiou:2010pv}
  C.~Athanasiou, P.~M.~Chesler, H.~Liu, D.~Nickel and K.~Rajagopal,
  ``Synchrotron radiation in strongly coupled conformal field theories,''
  Phys.\ Rev.\ D {\bf 81}, 126001 (2010)
  [Erratum-ibid.\ D {\bf 84}, 069901 (2011)]
  [arXiv:1001.3880 [hep-th]].
  %%CITATION = ARXIV:1001.3880;%%


  \bibitem{Semenoff:2001xp}
  G.~W.~Semenoff and K.~Zarembo,
  ``More exact predictions of SUSYM for string theory,''
  Nucl.\ Phys.\ B {\bf 616}, 34 (2001)
  [hep-th/0106015].
  %%CITATION = HEP-TH/0106015;%%.


 \bibitem{Gaiotto}
D. Gaiotto, private communication.


\bibitem{Basso:2011rs}
  B.~Basso,
  ``An exact slope for AdS/CFT,''
  arXiv:1109.3154 [hep-th].
  %%CITATION = ARXIV:1109.3154;%%


\bibitem{Nishioka:2008gz}
  T.~Nishioka and T.~Takayanagi,
  ``On Type IIA Penrose Limit and N=6 Chern-Simons Theories,''
  JHEP {\bf 0808}, 001 (2008)
  [arXiv:0806.3391 [hep-th]].

  %\cite{Gaiotto:2008cg}

\bibitem{Gaiotto:2008cg}
  D.~Gaiotto, S.~Giombi and X.~Yin,
  ``Spin Chains in N=6 Superconformal Chern-Simons-Matter Theory,''
  JHEP {\bf 0904}, 066 (2009)
  [arXiv:0806.4589 [hep-th]].
  %%CITATION = ARXIV:0806.4589;%%

%\cite{Grignani:2008is}

\bibitem{Grignani:2008is}
  G.~Grignani, T.~Harmark and M.~Orselli,
  ``The SU(2) x SU(2) sector in the string dual of N=6 superconformal Chern-Simons theory,''
  Nucl.\ Phys.\ B {\bf 810}, 115 (2009)
  [arXiv:0806.4959 [hep-th]].
  %%CITATION = ARXIV:0806.4959;%%

%\cite{Gromov:2008qe}

\bibitem{Gromov:2008qe}
  N.~Gromov and P.~Vieira,
  ``The all loop AdS4/CFT3 Bethe ansatz,''
  JHEP {\bf 0901}, 016 (2009)
  [arXiv:0807.0777 [hep-th]].
  %%CITATION = ARXIV:0807.0777;%%
  
%\cite{Fiol:2012sg}
\end{thebibliography}
\end{document}